\documentclass[sigconf,nonacm]{acmart}

\usepackage{graphicx}
\usepackage{svg}
\usepackage{caption}
\usepackage{subcaption}

\setcopyright{rightsretained}
\acmConference[SIGIR eCom'22]{ACM SIGIR Workshop on eCommerce}{July 15, 2022}{ Madrid, Spain}
\acmYear{2022}
\copyrightyear{2022}
\makeatletter \renewcommand\@formatdoi[1]{\ignorespaces}
\makeatother
\acmISBN{}

\begin{document}

\title{UNDR: User-Needs-Driven Ranking of Products in E-Commerce}

\author{Andrea Papenmeier}
\email{andrea.papenmeier@uni-due.de}
\affiliation{%
  \institution{University of Duisburg-Essen}
  \city{Cologne}
  \country{Germany}
}
\author{Daniel Hienert}
\email{daniel.hienert@gesis.org}
\affiliation{%
  \institution{GESIS – Leibniz Institute for the Social Sciences}
  \city{Cologne}
  \country{Germany}
}
\author{Firas Sabbah}
\email{firas.sabbah@uni-due.de}
\affiliation{%
  \institution{University of Duisburg-Essen}
  \city{Duisburg}
  \country{Germany}
}
\author{Norbert Fuhr}
\email{norbert.fuhr@uni-due.de}
\affiliation{%
  \institution{University of Duisburg-Essen}
  \city{Duisburg}
  \country{Germany}
}
\author{Dagmar Kern}
\email{dagmar.kern@gesis.org}
\affiliation{%
  \institution{GESIS – Leibniz Institute for the Social Sciences}
  \city{Cologne}
  \country{Germany}
}

\renewcommand{\shortauthors}{Papenmeier et al.}

\begin{abstract}
  Online retailers often offer a vast choice of products to their customers to filter and browse through. The order in which the products are listed depends on the ranking algorithm employed in the online shop. State-of-the-art ranking methods are complex and draw on many different information, e.g., user query and intent, product attributes, popularity, recency, reviews, or purchases. However, approaches that incorporate user-generated data such as click-through data, user ratings, or reviews disadvantage new products that have not yet been rated by customers. We therefore propose the \emph{User-Needs-Driven Ranking} (\emph{UNDR}) method that accounts for explicit customer needs by using facet popularity and facet value popularity. As a user-centered approach that does not rely on post-purchase ratings or reviews, our method bypasses the cold-start problem while still reflecting the needs of an average customer. In two preliminary user studies, we compare our ranking method with a \emph{rating-based ranking} baseline. Our findings show that our proposed approach generates a ranking that fits current customer needs significantly better than the baseline. However, a more fine-grained usage-specific ranking did not further improve the ranking.
\end{abstract}



\maketitle

\section{Introduction}
Ranking methods for product search are very complex. Aiming to combine business goals with user needs~\cite{derakhshan2020product, tsagkias2021}, ranking methods usually combine a large number of features, such as product descriptions, user ratings, reviews, and behavioral data like clicks or purchases \cite{chaabna2015designing, najmi2015capra, santu2017, sorokina2016}. From reviews, ranking algorithms can extract for example information about the importance of individual product attributes for users and subsequently account for those user needs in the ranking score~\cite{voigt2012weighted, zha2014product}.

However, for ``cold'' products~\cite{pourgholamali2016mining}, i.e., products that are new or have not yet been purchased and evaluated, user-generated data is not available. Those products pose a problem for ranking methods that aim to reflect user needs based on user-generated data. Moreover, user ratings, reviews and click-through data carry implicit information about user needs and need to be carefully cleaned and preprocessed. To overcome the problem of cold products, literature suggested estimating missing data based on reviews from similar products~\cite{gupta2020, missault2021addressing} or using alternative information sources such as social media to gather insights about cold products~\cite{pourgholamali2016mining, zhao2016connecting}. 

\begin{figure}[ht]
    \centering
    \includegraphics[width=0.9\linewidth]{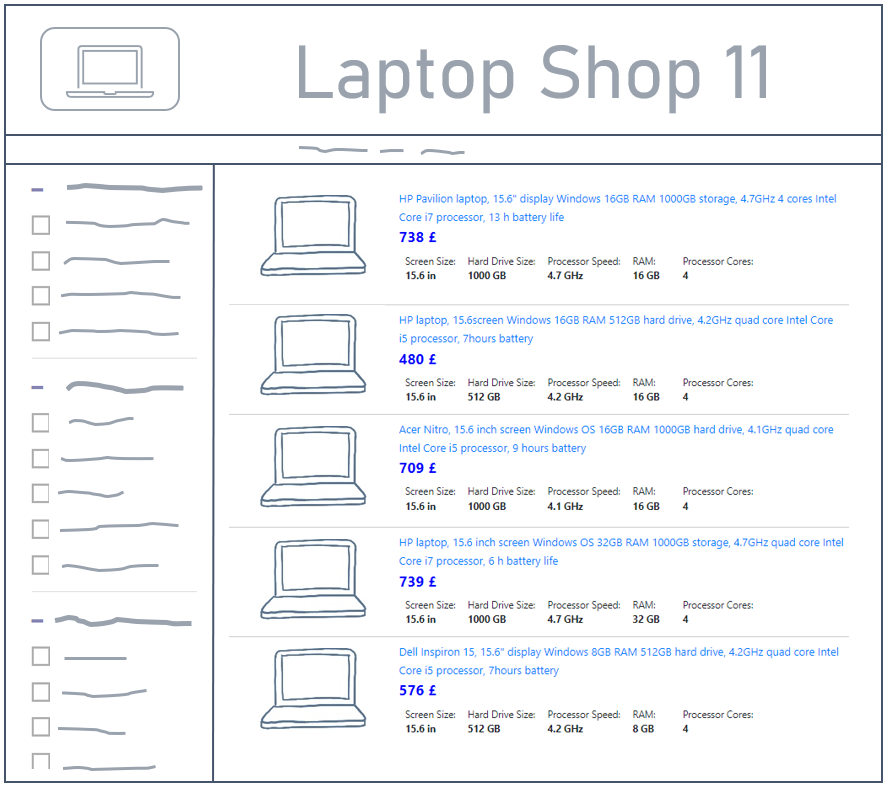}
    \caption{Screenshot of a fictive laptop shop with the \emph{UNDR} method.}
    \label{fig:study2_screenshotUNDR}
\end{figure}

To provide an alternative method for gathering information about attribute-level user needs, we propose the \emph{User-Needs-Driven Ranking} (\emph{UNDR}) score that accounts for current user needs by 
\begin{enumerate}
    \item Gathering explicit, structured information about current user needs for a product category by collecting facet selections of users.
    \item Driven by the user needs, calculating the average popularity of each facet and its values.
    \item Using the popularity weights to assign popularity scores to each product and rank them accordingly. 
\end{enumerate}
To evaluate the \emph{UNDR} method, we collected the multi-faceted user needs for a new laptop in a crowdsourcing experiment with N~=~277 participants. We calculated the popularity weights for laptop attributes and their values, representing the average needs of all users. Subsequently, we computed the product ranking with the \emph{UNDR} method and compared it to a \emph{rating-based ranking} method in two user studies. In the first user study with N~=~59 participants, participants were confronted with screenshots of fictive laptop shops displaying the top 5 ranked laptops of each method (see Figure \ref{fig:study2_screenshotUNDR}). Participants assessed how well the offered laptops fit their needs as well as which store they would like to visit first. To explore the potential of more distinct user profiles, we conducted a second user study with N~=~144 participants. We compared a ``basic'' user profile, i.e., popularity weights calculated from basic laptop users, and an ``advanced'' user profile, i.e., popularity weights based on advanced laptop users, with the baseline. 

Our findings show that the top 5 laptops of the \emph{UNDR} method are perceived to better fit the user needs of our participants compared to the \emph{rating-based} baseline. Consistently, significantly more participants decided to first visit the \emph{UNDR} shop. However, the second user study showed that offering a more specialized ranking adapted to a user profile (basic, advanced) does not further improve how well the products fit participants' needs.

With our approach, we contribute a ranking method that reflects current user needs and can be deployed by shops without collecting post-purchase generated user data (therefore avoiding the cold-start problem). With the user evaluations, we further contribute empirical evidence in support of the proposed ranking method.

\section{Related Work}
Online shops facilitate the product discovery and purchase of customers in e-commerce. Compared to information discovery in web search, product search brings up unique challenges, such as integrating user needs with business needs~\cite{derakhshan2020product, tsagkias2021}, new performance evaluation criteria~\cite{wu2018clicks}, and data and data annotation challenges~\cite{tsagkias2021, wu2018clicks}. In online product search, search systems are confronted with users' complex, multi-faceted information needs~\cite{ben2008beyond, vandic2013facet, zha2014product}. That is, users have preferences for multiple product attributes, which is often accounted for with providing facets in the search interface~\cite{ben2008beyond, hearst2002finding}. While some product aspects are highly important to customers, others are less relevant~\cite{kern2018evaluation, voigt2012weighted, zha2014product}, e.g., when searching for a laptop, a user might find the price and brand very important but the processor brand less important. Besides product attributes, social attributes like the average product rating influence customer's purchase decision~\cite{li2016predicting, poston2005effective}.

Information about the importance of product aspects can be used to improve the \textbf{ranking of products}, i.e., the order in which the products are sorted and shown to the users. Often, users only consider the first products in the result list~\cite{derakhshan2020product}, making the ranking method one of the most crucial and complex issues in this field \cite{tsagkias2021}. Previous literature suggested either asking users to input and control their individual levels of importance of product aspects~\cite{kern2018evaluation}, or extracting the importance levels from user-generated product content such as user reviews~\cite{voigt2012weighted, zha2014product}. Taking a binary approach in which users can select important features (rather than weighting them), Sabbah et al.~\cite{sabbah2021} showed that aspect-level sentiment information from reviews can support ranking performance. Other works have mined review texts and user ratings (e.g., the 5-star rating that users assign to a product) to extract information for individual product features to improve the ranking~\cite{chaabna2015designing, najmi2015capra}. Besides user-generated reviews and ratings, other user data can be used to improve the ranking: Wu et al.~\cite{wu2018clicks} leveraged click-through data and information about purchases to optimize product ranking, and Derakhshan et al.~\cite{derakhshan2020product} used the ``consideration set'' of users (i.e., the set of products that users have encountered during their search) in their ranking model. Overall, ranking algorithms are often complex models that combine a number of features both from users and from sellers, such as user queries, click-through data, add-to-carts, revenue information, or order rates \cite{santu2017}. For a more complete overview over ranking algorithms used in e-commerce, the reader is referred to Najib et al.~\cite{najib2019systematic} and Santu et al.~\cite{santu2017}. 

Besides comparing several ranking methods, Santu et al.~\cite{santu2017} also derived a set of challenges for the field of product search. One is the ``presence of uncertain features'' which especially applies for new products: ``\textbf{Cold products}''~\cite{pourgholamali2016mining}, i.e., new products that were just released in an online shop, initially do not have user reviews, ratings, or click-through data, hence introducing uncertain features. To overcome the cold-start problem, literature proposes several approaches. One line of research estimates missing information based on attributes and information a new product shares with other products in the same or a similar domain (e.g., in \cite{gupta2020, missault2021addressing, pourgholamali2016mining}). Similarly, Zhao et al.~\cite{zhao2016connecting} aggregated data from social media and use them as substitutions for missing reviews of cold products in product recommendation, while Pourgholamali~\cite{pourgholamali2016mining} collected product data across several additional sources to fill for missing reviews and ratings. Other works use continuous data collection during search, for example, Bi et al. \cite{bi2019leverage} recorded a user's clicking behavior on the first result page for re-ranking later result pages within a single search session.

User behavior data is also used in the related field of recommendation systems to improve \textbf{personalized} product recommendations~\cite{ding2020implicit, zhou2018micro} or personalized product rankings~\cite{zhang2017personalized}. Additionally, structured knowledge about the landscape of products, i.e., hierarchical relations between product categories, can be deployed to improve recommendations~\cite{yu2020personalized} and to build user profiles~\cite{gu2020hierarchical}.


\section{UNDR: User-Needs-Driven Ranking}
In this section, we introduce our \emph{User-Needs-Driven Ranking} (\emph{UNDR}) method which assigns user-centered ranking scores to products without the need for user-generated post-purchase data such as ratings or reviews.

While previous ranking and recommendation methods suggest harnessing reviews and ratings~\cite{chaabna2015designing, najmi2015capra, sabbah2021, zha2014product}, we propose gathering information about users' facet selection behavior as indicator for user needs in a product browsing context. This data can either be collected independent of an online shop with controlled user surveys or, if an online shop with facets is already in use, by logging users' facet behavior directly. In contrast to ratings and reviews, a facet-based approach does not require actual purchases and an extra effort of rating and reviewing the product. From the facet selection behavior, we can derive (1) how popular a facet or product attribute is for users, and (2) how popular a specific facet value or attribute value is for users. We assume that:
\begin{itemize}
    \item[\textbf{A1}] The more often a facet is selected, the more important the attribute is to users.
    \item[\textbf{A2}] The more often a facet value is selected, the stronger this value represents the current average user need.
\end{itemize}
For a set of facets $F$, we can then compute a user-needs-driven ranking score for a product as follows:
\begin{equation}
    \text{UNDR}_{\rm score} = \sum_{f \in F} w_f\cdot w_{f_v}
\end{equation}
where $f$ is a facet or product attribute in $F$, $w_{f}$ the popularity weight of a facet (computed as the number of users that used $f$ divided by the total number of observed users), and $w_{f_v}$ the popularity weight of a facet value (computed as the number of users that selected value $f_v$ divided by the number of users that used $f$).

To evaluate our proposed ranking approach, we explore the use case of laptop shops. Laptops are regularly used by many people (simplifying the recruitment of target users), are usually described by multiple attributes (hence representing a multi-faceted, complex need), and are already sold online so that collecting a dataset with user-generated information as a baseline is feasible. In the following sections, we describe how the \emph{UNDR} method can be used in the context of an online laptop shop.

\subsection{Collecting User Needs}
\label{ssec:userNeedsCollection}
\begin{figure}[ht]
    \centering
    \includegraphics[width=\linewidth]{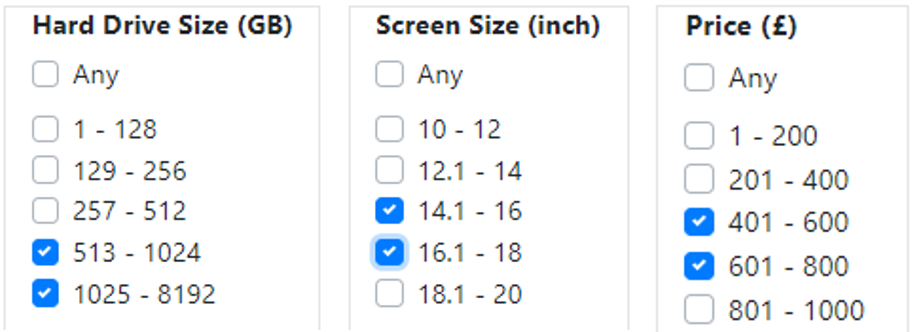}
    \caption{Example of facet value selections for laptop attributes ``Hard Drive Size'', ``Screen Size'', and ``Price''.}
    \label{fig:study1_filter1}
\end{figure}

\begin{figure}[ht]
    \centering
    \includesvg[width=\linewidth]{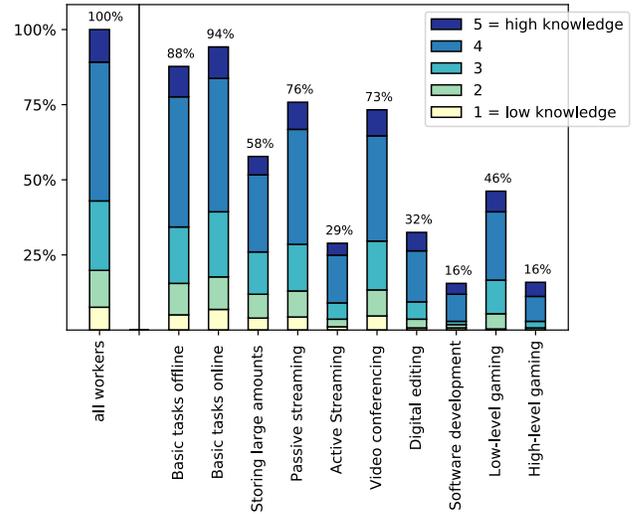}
    \caption{Participants laptop usage habits per task (multiple choice) in percent, divided into self-reported levels of domain knowledge about laptops.}
    \label{fig:study1_tasks}
\end{figure}

The \emph{UNDR} method uses information about the popularity of facets and their values. Data of facet selections can be collected either from usage logs of an existing online shop, or by a structured user survey. In our experiment, we set up a crowdsourcing task to collect information about facet usage. We asked workers to first answer questions about their demographic background (age, gender, domain knowledge about laptops) and for which tasks they usually use their laptop (multiple choice of ten common tasks, see Figure~\ref{fig:study1_tasks}). Workers then read a scenario (``Imagine your computer broke down. Now, you are planning to buy a new laptop.'') and were asked to make a facet selection for ten common laptop attributes (price, RAM size, operating system, brand, hard drive size, screen size, CPU cores, CPU speed, CPU brand, battery life). See Figure~\ref{fig:study1_filter1} for an exemplary facet selection. We asked workers to select the ``any'' option, if the attribute was not important to them. 

We recruited N~=~304 crowd workers on \textit{Prolific}\footnote{\url{https://www.prolific.co}}, who had to be English native speakers currently residing in the United Kingdom without literacy difficulties. We restricted the participation to a single country to reduce effects of the current economic situation and accompanying variance in the conception of laptop prices. We excluded 14 responses due to low quality and 13 responses because workers did not own a laptop, which could affect their frame of reference. The final sample consisted of 277 workers (160 female, 110 male, 3 non-binary, 4 prefer not to say) with an average age of M~=~36.5 years (SD~=~11.4 years) and a medium domain knowledge about laptops (on a scale of 1 = ``low knowledge about laptops'' to 5 = ``high knowledge'', M~=~3.6, SD~=~1.0). Figure~\ref{fig:study1_tasks} shows the distribution of domain knowledge levels per laptop usage task. Some purposes, like basic tasks, streaming, and video conferencing are done by workers across all domain knowledge levels, whereas more advanced tasks (e.g., software development, high-level gaming) are mostly done by workers with high knowledge of laptops.

\subsection{UNDR Popularity Weights}
\label{ssec:UNDR_weights}
Based on the responses, we calculated the overall popularity weight of each attribute as the percentage of users who selected a specific value, i.e., percentage of users who did not select the ``any'' option. Table~\ref{tab:study1_popularityWeights} lists the attribute-level popularity weights to give an understanding of the relative popularity of the ten laptop attributes. For example, for the attribute ``Screen Size'', 41 users (15\%) selected the ``any'' option. The overall popularity weight of ``Screen Size'' is therefore $w_{f} = 0.85$. Furthermore, out of the $277 - 41 = 236$ users who selected specific screen size values, 153 users (40\%) selected the value ``14.1 - 16'' inches. Consequently, the value-level popularity weight of ``14.1 - 16'' in ``Screen Size'' is $w_{f_v} = 0.40$. A laptop with a screen size of 14.9 would therefore receive a score of $0.85 \cdot 0.40 = \mathbf{0.34}$ for the screen size attribute. Screens smaller than 12 inches are less popular and received a value-level popularity of $w_{f_v} = 0.03$ in our experiment. A laptop with an 11 inches screen would therefore be assigned a score of $0.85 \cdot 0.03 = \mathbf{0.026}$ for the screen size attribute.

\begin{table}[h!]
    \begin{center}
        \begin{tabular}{ c|c c } 
             attribute & ``any'' count & overall \\ 
              & out of 277 & popularity weight \\
             \midrule
             Price              & 24 & 0.91 \\ 
             Brand              & 77 & 0.72 \\ 
             Operating system   & 35 & 0.87 \\
             Screen size        & 41 & 0.85 \\ 
             Hard drive size    & 55 & 0.80 \\ 
             RAM size           & 33 & 0.88 \\  
             CPU cores          & 126 & 0.55 \\ 
             CPU speed          & 90 & 0.68 \\ 
             CPU brand          & 141 & 0.49 \\ 
             Battery life       & 38 & 0.86 \\ 
        \end{tabular}
        \caption{Overall popularity weights per laptop attribute.}
        \label{tab:study1_popularityWeights}
    \end{center}
\end{table}

\subsection{Product Dataset Collection}
We collected a dataset of 1,445 laptops from \textit{Amazon}\footnote{\url{https://www.amazon.com}} in December 2021, containing product information such as the technical descriptions, prices, and user-generated data such as ratings and reviews. Prices were collected in US dollars and converted to pound sterling using the exchange rate of those days of 0.75. We discarded data points that did not carry information about all ten common laptop attributes (price, RAM size, operating system, brand, hard drive size, screen size, CPU cores, CPU speed, CPU brand, battery life) and duplicates. Furthermore, to allow for comparison with a rating-based baseline, we only kept data points with at least ten customer ratings and calculated for each laptop the average rating score. After the reduction, the final dataset contains 182 laptops. Using the popularity weights for attributes and their values (see Section~\ref{ssec:UNDR_weights}), we then calculated the \emph{UNDR} score for each laptop and ranked the laptops, starting with the laptop with the highest \emph{UNDR} score at the first rank.

\section{User Study 1: UNDR vs. Baseline}
\label{sec:user1}
The \emph{UNDR} method aims to be a user-centered ranking approach without needing user-generated post-purchase data such as ratings or reviews. To evaluate whether our method can substitute or even improve on common product popularity signals such as star ratings, we designed a user study answering the following research question:
\begin{itemize}
    \item[\textbf{RQ1}] How well does the \emph{UNDR} method perform compared to a \emph{rating-based ranking} method in the eyes of the users?
\end{itemize}
In this preliminary study, we chose the ``sort-by-average customer rating''-function as our baseline because this function is well-known to users and commonly offered by most online-shops as a sorting feature. We compared the two rankings (\emph{UNDR}, \emph{rating-based ranking}) using a within-subject design. Here, we focused on the ``first impression'' of rankings and investigated whether users are more drawn towards a shop using the \emph{UNDR} method for ranking as opposed to a shop using a \emph{rating-based ranking}. 
Before investing effort and time into the development of a fully interactive prototype, we decided to start our research with a simplistic, preliminary user study. We used static screenshots (showing the top 5 results of each ranking method) to collect initial insights into the potential of the \emph{UNDR} score.

\subsection{Task and Procedure}
\begin{figure*}[t]
     \centering
     \begin{subfigure}[b]{0.49\textwidth}
         \centering
         \includegraphics[width=\linewidth]{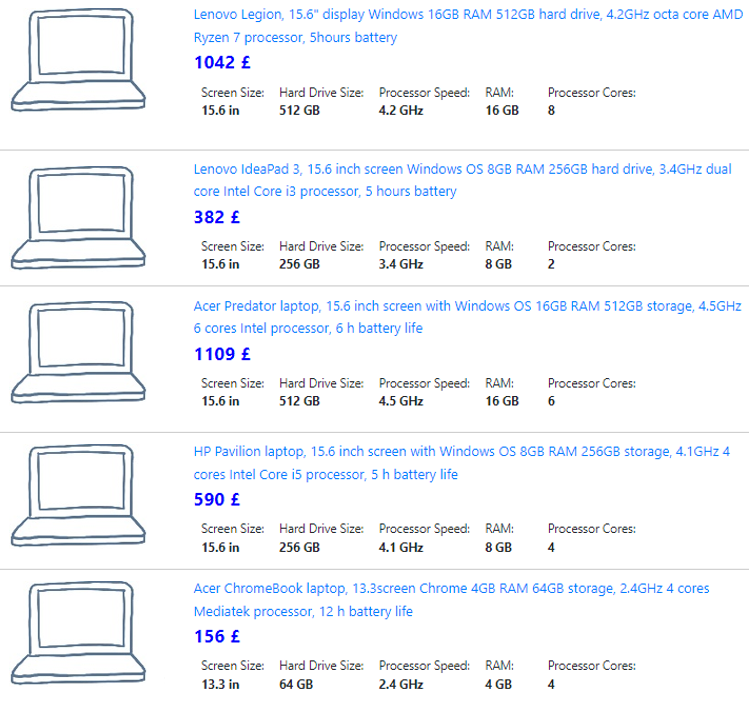}
         \caption{Laptop list with the \emph{rating-based ranking} method.}
         \label{fig:study2_top5Baseline}
     \end{subfigure}
     \hfill
     \begin{subfigure}[b]{0.49\textwidth}
         \centering
         \includegraphics[width=\linewidth]{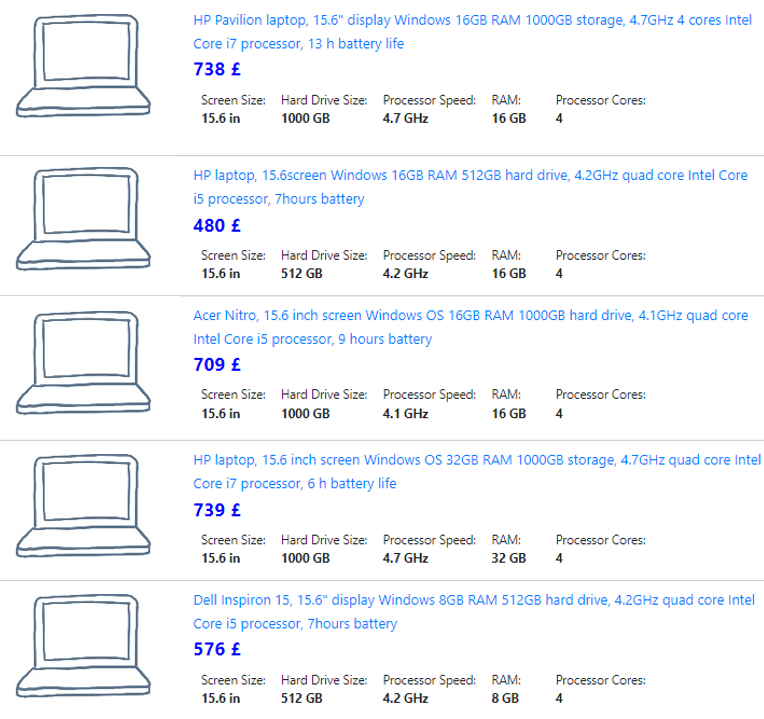}
         \caption{Laptop list with the \emph{UNDR} method.}
         \label{fig:study2_top5UNDR}
     \end{subfigure}
     \caption{Screenshot of the top 5 results of both fictive laptop shops.}
     \label{fig:study2_screenshots}
\end{figure*}

We asked participants of our user study to first give informed consent and answer some demographic questions (age, gender, domain knowledge about laptops). Subsequently, they read the same scenario description as the crowdworkers (see Section~\ref{ssec:userNeedsCollection}) that prompted them to imagine needing a new laptop that fits their current needs. Furthermore, the scenario told them that they find two online shops on the internet. Participants then saw two static screenshots: One of the \emph{rating-based} shop and one of the \emph{UNDR} shop. The exact result lists presented to the participants are shown in Figure~\ref{fig:study2_top5Baseline} (baseline) and Figure~\ref{fig:study2_top5UNDR} (\emph{UNDR}). We constructed the screenshots such that they resemble the common online shop structure with shop branding at the top, a facet column on the left, and a result list in the center (see Figure~\ref{fig:study2_screenshotUNDR}). We replaced the facet column with scribbles and product images with a generic image to reduce the effect of visual confounders~\cite{wang2016impact}. All laptops in the result list display the same information and have a standardized title (model, brand, screen size, operating system, RAM size, storage size, processor information, battery life). To simulate shops with ``cold'' products, i.e., products without ratings or reviews, we did not display the user ratings -- the ratings were only used for determining the ranking of the \emph{rating-based} shop. The order in which the screenshots were presented was randomised. For each shop, participants were asked to indicate how well the shop fits their needs and how likely it is that they would visit this store. Finally, participants selected which shop they would visit first and described the reasons for their decision in an open text box.

\subsection{Measures and Analysis}
\label{ssec:user1_measures}
To investigate how the two ranking methods perform in the eyes of the users, we analyzed three performance indicators:

\begin{enumerate}
    \item \textbf{Fitness}: Rating of agreement with the statement ``The laptops of shop \textit{X} fit my needs.'' on a scale of 1 (``strongly disagree'') to 5 (``strongly agree''). We tested for significant differences using the Wilcoxon signed-rank test for paired samples at $\alpha = 0.05$.
    \item \textbf{Visit likelihood}: Rating of agreement with the statement ``I would like to visit the website of shop \textit{X}.'' on a scale of 1 (``strongly disagree'') to 5 (``strongly agree''). We tested for significance with the Wilcoxon signed-rank test at $\alpha = 0.05$.
    \item \textbf{Shop selection}: Number of times a shop was selected in the question ``Which of the two laptop shops would you visit first?''. We determined significant differences using the cumulative distribution function of the binomial distribution, testing whether the observed distribution is different from a 50\%-50\% distribution at $\alpha = 0.05$.
\end{enumerate}

Furthermore, we added a qualitative measure to gain additional insights into users' decision-making processes:

\begin{enumerate}
    \item [(4)] \textbf{Shop selection reason}: Open text answer to the question ``Why do you think this shop best fits your needs?''. We analyzed the data with a qualitative coding and clustering approach with one annotator.
\end{enumerate}

\subsection{Participants}
\label{ssec:user1_participants}
Similar to the recruitment in Section~\ref{ssec:userNeedsCollection}, we invited N~=~60 participants on \textit{Prolific} to take part in our user study. We applied the same prescreening aspects as before (English native speakers, UK residents, no literacy difficulties) to reduce effects of cultural or economic confounders. Within the N~=~59 valid responses (one discarded due to low quality), age distribution (M~=~36.6~years, SD~=~13.4~years) was similar to the crowdsourcing experiment and the gender distribution was balanced (29~female, 30~male, 0~non-binary, 0~prefer not to say). Participants had, again, a medium average domain knowledge about laptops (on a scale of 1 to 5, M~=~3.2, SD~=~1.1).

\subsection{Results}
\begin{figure}[b]
    \centering
    \includesvg[width=\linewidth]{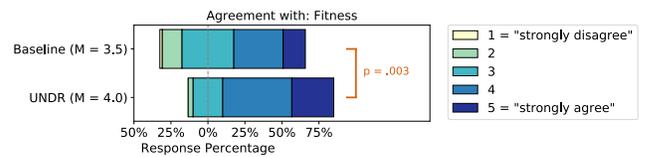}
    \caption{Likert plot of fitness measure for both ranking interfaces and Wilcoxon's test result, p-value corrected with Bonferroni method.}
    \label{fig:study2_fitness}
\end{figure}

To investigate how well the \emph{UNDR} method performs compared to a \emph{rating-based ranking} method in the eyes of the users (\textbf{RQ1}), we first looked at how well both ranking methods fit users' needs. Figure~\ref{fig:study2_fitness} depicts the distribution of agreement levels with the statement ``The laptops of shop X fit my needs'' in the baseline shop (M~=~3.5, SD~=~1.0) and \emph{UNDR} shop (M~=~4.0, SD~=~0.8). The difference in means is significant (U~=~172, p~=~.003), showing that participants found the offer of the \emph{UNDR} shop to better fit their needs than the laptops of the \emph{rating-based} baseline.

\begin{figure}[b]
    \centering
    \includesvg[width=\linewidth]{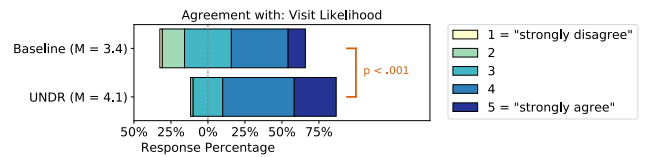}
    \caption{Likert plot of visit likelihood measure for both ranking interfaces and Wilcoxon's test result, p-value corrected with Bonferroni method.}
    \label{fig:study2_visitLikelihood}
\end{figure}

We further looked at the visit likelihood as a second measure of how participants perceive and assess the rankings. On average, participants reported a significantly higher visit likelihood (U~=~94, p~<~.001) for the \emph{UNDR} shop (M~=~4.1, SD~=~0.7) than for the baseline shop (M~=~3.4, SD~=~0.9). Figure~\ref{fig:study2_visitLikelihood} visualizes participants' responses in the visit likelihood measure.

At the end of the experiment, participants had to make a decision about which of the two stores to visit first. 39 participants (66\%) decided for the \emph{UNDR} shop, whereas 20 participants (34\%) selected the baseline shop. This distribution differs significantly (p~=~.009) from an expected 50\%-50\% distribution of both shops would draw an equal number of visitors. Participants also explained their decision in an open answer. 37 participants mentioned the price of the selected shop to either better fit their needs or to be closer to the price range they would expect for a laptop at the time. Most of the times, participants set the observed prices in relation to the hardware specifications: \textit{``They seem better value for money''} or \textit{``The pricing of shop 4 seems to be higher for the spec''}. Furthermore, participants often mention specific laptop attributes that seem to be important for them: hard drive size (15), RAM size (5), CPU (4), brand (4), battery life (3), and screen size (1). Similarly, others do not describe specific attributes, but note that the specifications fit their specific usage goal such as \textit{``help with daily tasks''}, \textit{``for personal use''}, while 15 say it fits their needs in general: \textit{``Better selection for my needs''} or \textit{``match my requirements''}. Although participants report foremost price and fitness for their (individual) needs as the deciding factors, more subtle factors seem to effect the decision as well. While 6 participants liked the broader price range in the baseline shop (\textit{``Broad range of price points''}), others take the narrow price range in the \emph{UNDR} shop as a sign of a well-curated laptop offer: \textit{``the prices and descriptions are similar which suggest they are of a certain standard''} and \textit{`` It looks like they are more evenly priced, which means they are likely acceptable quality''}.

\section{User Study 2: Profiling}
\label{sec:user2}
The \emph{UNDR} method can potentially provide a ranking tailored to specific user groups with different usage habits, depending on whose data the popularity weights are based on. In the first user study (Section~\ref{sec:user1}), we used a ranking based on the data of all crowd workers. However, while some crowd workers used their laptops only for ``basic'' tasks, others, more knowledgeable crowd workers, also had ``advanced'' usage habits (see Figure~\ref{fig:study1_tasks}). We therefore set up a second user study to answer the following research questions:
\begin{itemize}
    \item[\textbf{RQ2}] Do user-group-specific (basic, advanced) \emph{UNDR} result lists better fit the users' needs than the general \emph{UNDR} result list or the \emph{rating-based} baseline?
    \item[\textbf{RQ3}] To what extent does the classification into an incorrect user profile harm the user's view of the ranking performance?
\end{itemize}
Using the data from the crowdsourcing experiment, we calculated a \emph{basic UNDR} laptop order based on the data of the 83 workers with only basic usage habits (no digital editing, no software engineering, no high-level gaming). Equally, we computed an \emph{advanced UNDR} order on the data of the 194 workers with advanced usage habits (at least one of: digital editing, software engineering, high-level gaming). 
We used a study design similar to the setup in the first user study: We compared \emph{UNDR} shops with the \emph{rating-based} baseline shop in a within-subject experiment. Moreover, we made sure to gather at least 30 valid responses for any of the following groups: The correct profile classifications (1) basic user - \emph{basic UNDR} shop and (2) advanced user - \emph{advanced UNDR}, and the incorrect profile classifications (3) basic user - \emph{advanced UNDR} shop and (4) advanced user - \emph{basic UNDR}.

\subsection{Task and Procedure}
For comparability, we kept the task description, procedure and questions of the first user study and only exchanged the screenshots of the \emph{UNDR} shops.

\subsection{Measures and Analysis}
In this experiment, we focused only on the quantitative measures of \textbf{fitness} to needs as described in Section~\ref{ssec:user1_measures}. For comparison with the baseline (within-subject comparison), we again used the Wilcoxon signed-rank test for paired samples. To compare the fitness of two \emph{UNDR} shops (e.g., general \emph{UNDR} from the first user study vs \emph{basic UNDR} or \emph{advanced UNDR}), we used the Mann-Whitney U test. All significance tests are evaluated at a significance level of $\alpha = 0.05$.

\subsection{Participants}
We again used the recruitment procedure and prescreening factors described in Section~\ref{ssec:user1_participants} for this second user study. We stopped recruiting once we had at least 30 participants for each user - profile combination, leading to N~=~144 valid responses in total (n~=~36 in group (1), n~=~38 in group (2), n~=~33 in group (3), n~=~37 in group (4)). The sample group was approximately gender-balanced (76~female, 63~male, 3~non-binary, 2~prefer not to say) and on average slightly older (M~=~40.5~years, SD~=~14.5~years) than the participants of the first user study (but not significantly, Mann-Whitney U~=~3754, p~=~.070). The average domain knowledge about laptops was M~=~3.2 (SD~=~1.1).

\subsection{Results}
\begin{figure*}[ht]
    \centering
    \includesvg[width=\linewidth]{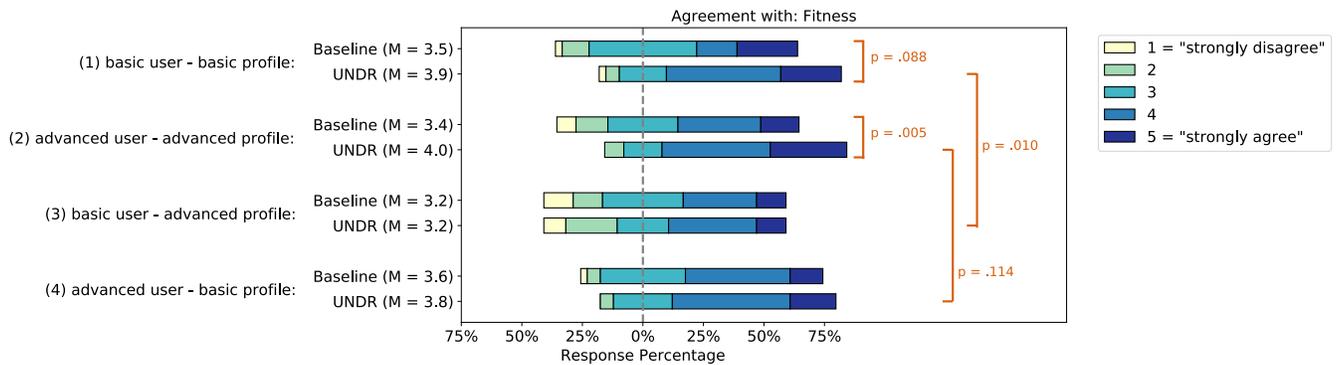}
    \caption{Likert plot of fitness measure in the second user study, grouped by actual usage habit (user group) and profile shown (basic, advanced). Each group assessed both the shop using \emph{UNDR} and the shop using the \emph{rating-based} baseline.}
    \label{fig:study3_fitness}
\end{figure*}

We first analyzed how the user-group-specific \emph{UNDR} shops are perceived by the respective user groups (\textbf{RQ2}). Figure~\ref{fig:study3_fitness} gives an overview of how well the shops fitted users' needs in all four groups. We did not find a significant difference between the \emph{basic UNDR} shop and the baseline for basic users in group (1) at $\alpha = 0.05$ (U~=~66, p~=~.088). For advanced users, the \emph{advanced UNDR} shop was a better fit than the baseline shop (U~=~62, p~=~.005). However, comparing how well the profiled \emph{UNDR} shops matched the needs of their users with how well the general \emph{UNDR} shop from the first study, we did not find a significant difference (U~=~2183, p~=~.431). Therefore, we cannot conclude that the profiling brings an advantage over the general \emph{UNDR} result list.

Besides the general potential for improvement with usage profiles, we investigated whether an incorrect profiling can harm how well the results fit users' needs (\textbf{RQ3}). That is, what happens if a basic user is mistaken as advanced user and subsequently sees the \emph{advanced UNDR} shop? 
On the one hand, for the basic user group (3), the \emph{advanced UNDR} shop is perceived to fit significantly less to the user needs than the \emph{basic UNDR} shop (U~=~409, p~=~.010). That is, mistaking a basic user for an advanced user and subsequently showing the wrong \emph{UNDR} ranking can potentially harm the user experience. On the other hand, the advanced user group (4) did not find the offer of the \emph{basic UNDR} shop to be a worse fit for their needs (U~=596, p~=~.114).





\section{Discussion}
In this paper, we introduce the \emph{User-Needs-Driven Ranking} (\emph{UNDR}) and take a first step to evaluate its potential as a user-centered ranking in a product browsing context. Our goal was to find a ranking that accounts for current user needs without requiring user-generated post-purchase data like ratings or reviews. In a first user study in the laptop domain, we explored how the \emph{UNDR} method compares to a \emph{rating-based} baseline at a ``first impression'' stage, similar to a situation in which a user enters an online shop and is confronted with an initial ranking of the products. Our findings show that the \emph{UNDR} method brought about products that are more fitting to users' needs than the baseline. Participants also reported to more likely visit the \emph{UNDR} shop than the baseline shop, and more often chose to visit the \emph{UNDR} shop first, indicating that this shop provides a higher chance of success in their eyes. Although the first user study was only a starting point, we derive from it that the \emph{UNDR} method delivers acceptable initial results and is worth exploring further in follow-up experiments. The \emph{UNDR} method could be especially valuable for online shops that have little user-generated data, e.g., shops that just went online, shops with fewer visitors, or shops that have less customers in some product categories than in others. It could also provide an advantage in product domains with fast-changing user needs (e.g., bikes, for which new features such as e-mobility were added over time) because user needs can be collected in a fast and structured way via surveys. 

In the second user study, we investigated the potential of a more fine-grained, user-group-specific ranking. We could not find an indication that a basic laptop ranking and an advanced laptop ranking with the \emph{UNDR} method provides added value over a general user profile. For participants with basic usage habits, showing the wrong profile could even reduce the positive effects of the \emph{UNDR} method. In our experiments, we had a well-controlled task with well-controlled recruiting procedure, which might have led to participants with similar needs. It is also possible that the classification into ``basic'' and ``advanced'' is not a suitable grouping factor; however, our findings also show that the profiling was not worse than the baseline. In other settings, needs might be more varied: Distinguishing for example between ``vegetarians'' and ``meat eaters'' in recipe search or between ``electric vehicle'' and ``car with combustion engine'' in car search might have a stronger effect.

Considering both user studies, we conclude: (1) Our findings indicate that the \emph{UNDR} method is a user-centered ranking that outperforms a \emph{rating-based ranking} baseline in a ``first impression'' user study. (2) We do not find an indication that further profiling into usage-based profiles (basic, advanced) provides an additional improvement of how well the product offer fits the users' needs.

\medskip
Since we present insights from preliminary studies, our findings are subject to several limitations. First, we used screenshots of online shops in our experiments, which eliminates the possibility for interactivity. We do not know whether the improved performance persists when adding facets and interactive elements such as textual search. However, a productive prototype of an online shop and bigger datasets (to avoid empty result lists) would be needed for an interactive experiment. In future work, we will develop an interactive prototype and evaluate ranking performance (e.g., using our fitness measure) at the end of a full search session. However, the insights from our user study show that shops could improve their ``first impression'' by using the \emph{UNDR} method. 
Moreover, the \emph{UNDR} method promotes products that conform to the average user need of a specific user group. Defining the boundaries of that target group, i.e., inclusion and exclusion criteria, clustering of similar users, is still an open question. Online shops that want to deploy the \emph{UNDR} method should make an extensive target group analysis to avoid rankings that fit the data but not the end-user. 
Finally, in this preliminary evaluation, we have treated the \emph{UNDR} score as an isolated metric. State-of-the-art ranking algorithms, however, are often more complex and consider multiple information sources. In a next step, the potential of integrating the \emph{UNDR} score with existing methods should be explored, e.g., in learning-to-rank approaches, either as an additional feature or as a proxy for other popularity signals such as customer ratings for cold products.
Despite those limitations, our initial experiments showcase the applicability of the \emph{UNDR} method to the laptop domain and promising performance in preliminary user experiments. 

\medskip
The \emph{UNDR} method is not only applicable to the laptop domain. In theory, it can be used in other product domains in which users have complex, multi-faceted information needs. The \emph{UNDR} score represents how well a product matches user needs with respect to a set of attributes without making assumptions about the form (can be applied to both categorical or numerical facets) or content of an attribute. It is therefore highly flexible and adaptable. Future studies should investigate the performance of the \emph{UNDR} score in various product domains such as technical products, clothes or furniture -- domains that vary in number of attributes and number of values per attribute. However, to give a useful score, the \emph{UNDR} method should only be applied in product domains where some attribute values are clearly more popular than others. In the laptop domain, for example, a screen smaller than 12 inches has a low popularity (.03) while 14 - 16 inches is much more popular (.40), which promotes 14 inches laptops and demotes smaller laptops. If all screen sizes were equally popular, the \emph{UNDR} score would not provide distinguishing information. 

Moreover, while we collected users' current facet selection behavior in a crowdsourcing experiment, it would be possible to collect such data from logs of productive systems. Future studies should explore how to collect and clean facet selection data to be used for calculating the \emph{UNDR} popularity weights.

\section{Conclusion}
This paper introduced the \emph{User-Needs-Driven Ranking} (\emph{UNDR}) score, an approach that utilizes facet selection behavior to bypass the cold-start problem while still delivering a user-centered ranking. We presented two preliminary user studies to investigate the potential applicability of the \emph{UNDR} method. Comparing our proposed method with a \emph{rating-based ranking} showed that the \emph{UNDR} method better addresses current user needs. However, further profiling the ranking method into usage-group specific rankings did not bring added value. If the \emph{UNDR} method persists to bring similar or even better results than methods based on user-generated data (e.g., click-through data or reviews) in future user studies, facet selection data could be a valuable addition to state-of-the-art ranking algorithms. Especially online shops with little user data could then profit from including the \emph{UNDR} score in their ranking to provide a user-centered ranking to their customers.

\begin{acks}
This work was partly funded by the DFG, grant no. 388815326; the VACOS project at GESIS and the University of Duisburg-Essen.
\end{acks}

\bibliographystyle{ACM-Reference-Format}
\bibliography{ecom22-bib}


\end{document}